\documentclass{ws-procs975x65-michal}

\newcommand{\kms}{\;\mathrm{km} / \mathrm{s}}
\newcommand{\bmg}{\mathbf{g}}
\newcommand{\bmr}{\mathbf{r}}
\newcommand{\bmv}{\mathbf{v}}
\newcommand{\Omm}{\Omega_\mathrm{m}} 
\newcommand{\vLG}{\bmv_\mathrm{_{LG}}}
\newcommand{\gLG}{\bmg_\mathrm{_{LG}}}
\newcommand{\degr}{^{\circ}}
\newcommand{\LCDM}{$\Lambda$CDM}

\begin{document}

\title{\uppercase{Precision cosmology with the 2MASS clustering dipole}}

\author{MACIEJ BILICKI$^*$,\\MICHA{\L} CHODOROWSKI, WOJCIECH HELLWING, THOMAS JARRETT, GARY MAMON}
\address{$^*$Astronomy, Cosmology and Gravity Centre, Department of Astronomy,\\
University of Cape Town, Republic of South Africa\\
E-mail: maciek@ast.uct.ac.za}

\begin{abstract}
Comparison of peculiar velocities of galaxies with their gravitational accelerations (induced by the density field) is one of the methods to constrain the redshift distortion parameter $\beta = \Omm^{0.55} / b$, where $\Omm$ is the non-relativistic matter density parameter and $b$ is the linear bias. In particular, one can use the motion of the Local Group (LG) for that purpose. Its peculiar velocity is known from the dipole component of the cosmic microwave background, whereas its acceleration can be estimated with the use of an all-sky galaxy catalog, from the so-called clustering dipole. At the moment, the biggest dataset of that kind is the Two Micron All Sky Survey Extended Source Catalog (2MASS XSC) containing almost 1 million galaxies and complete up to ~300 Mpc/h. We applied 2MASS data to measure LG acceleration and used two methods to estimate the beta parameter. Both of them yield $\beta\simeq0.4$ with an error of several per cent, which is the most precise determination of this parameter from the clustering dipole to date. 
\end{abstract}


\bodymatter

\bigskip

Motion of the Local Group (LG) of galaxies through the Universe probes the density parameter $\Omm$ via the linear-theory relation $\vLG=\beta\,\gLG$, where $\bmv$ and $\bmg$ stand respectively for the peculiar velocity and acceleration, and $\beta\equiv\Omm^{0.55}/b$, with $b$ being the linear bias. The velocity of the LG is known from the observed dipole anisotropy of the CMB temperature distribution\cite{Hinsh09} and equals to $\vLG=622\pm35\kms$. Its acceleration can be calculated with the use of an all-sky galaxy catalog (the \textit{clustering dipole}). In particular, as both the flux and the gravity fall off as distance squared, we can estimate it from the so-called \textit{flux dipole}:
\begin{equation}\label{gLG}
\gLG\sim\frac{H_0}{\rho_L}\sum_i S_i\hat{\bmr}_i
\end{equation}
where $S_i=L_i\slash 4\pi r_i^2$ is the flux received from the $i$-th galaxy with $L_i$ its absolute luminosity and $\rho_L$ is the universal luminosity density. 

Galaxy catalogs never reach down to zero flux and if we truncate the sum in Eq.~\eqref{gLG} at some magnitude limit, the clustering dipole thus estimated is a biased estimator of the actual peculiar acceleration of the LG. This is one of the reasons why we \emph{cannot} simply equate the clustering dipole and the velocity measured from the CMB to estimate the $\beta$ parameter. A more sophisticated approach is needed and we have developed two such methods\cite{BCJM11,TezaMB,BCH12}, applying them to the data from the Two Micron All Sky Survey.

The Two Micron All Sky Survey (2MASS\cite{Skr06}) is the first near-infrared ($JHK_s$ bands) survey of the whole sky and the 2MASS Extended Source Catalog (XSC\cite{Jar00}) is complete for sources brighter than \mbox{$K_s\simeq13.6$}~mag. The near-infrared flux is particularly useful for the purpose of large-scale structure studies as it samples the old stellar population, and hence the bulk of stellar mass, and it is minimally affected by dust in the Galactic plane\cite{Jar04}. An additional advantage of using 2MASS data is the global photometric uniformity of the catalog, and no bias nor offset between the photometry or astrometry obtained with the two telescopes used for the survey.

\begin{figure}
\begin{center}
\psfig{file=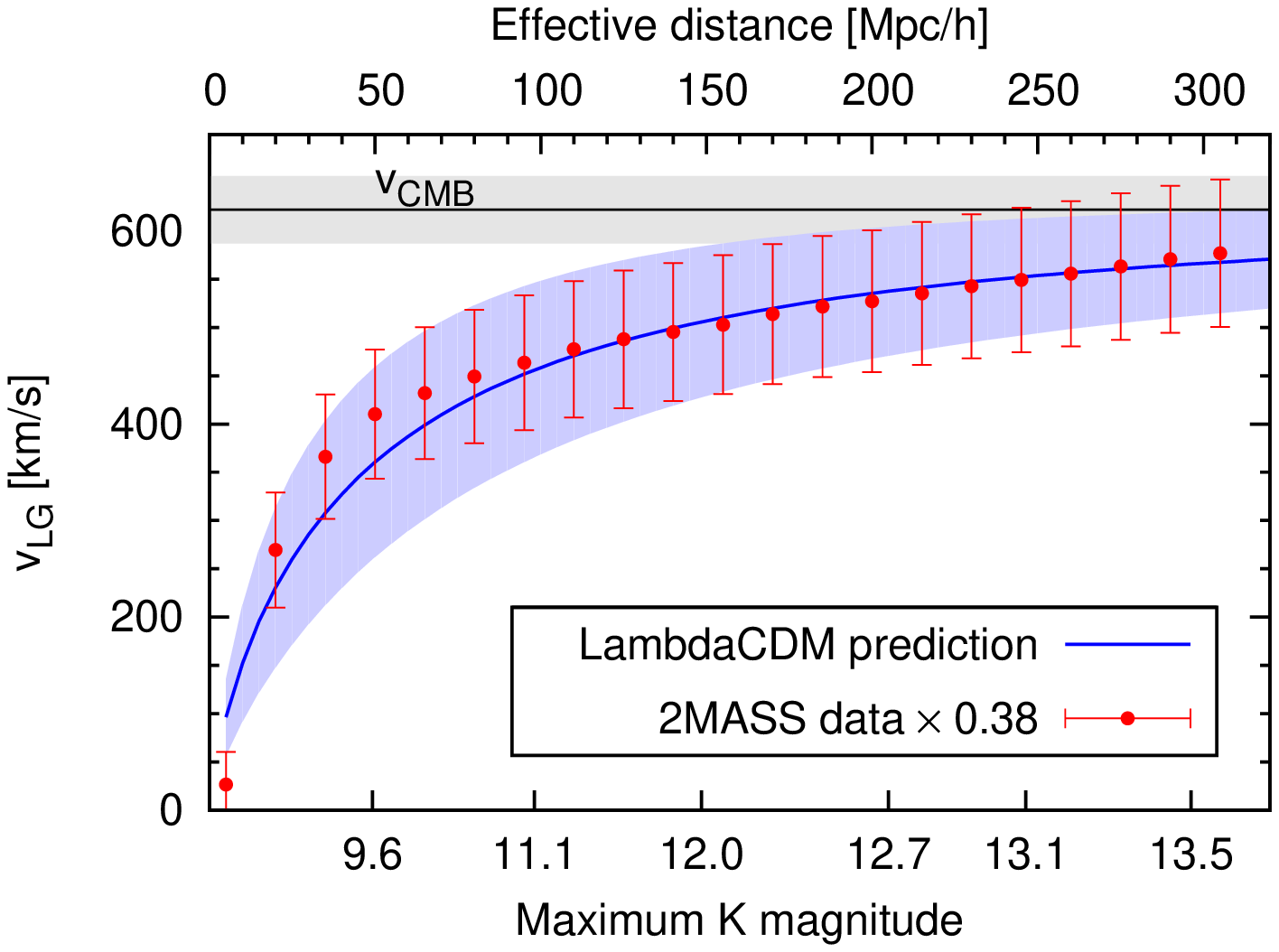,angle=-90,width=0.439\textwidth}
\psfig{file=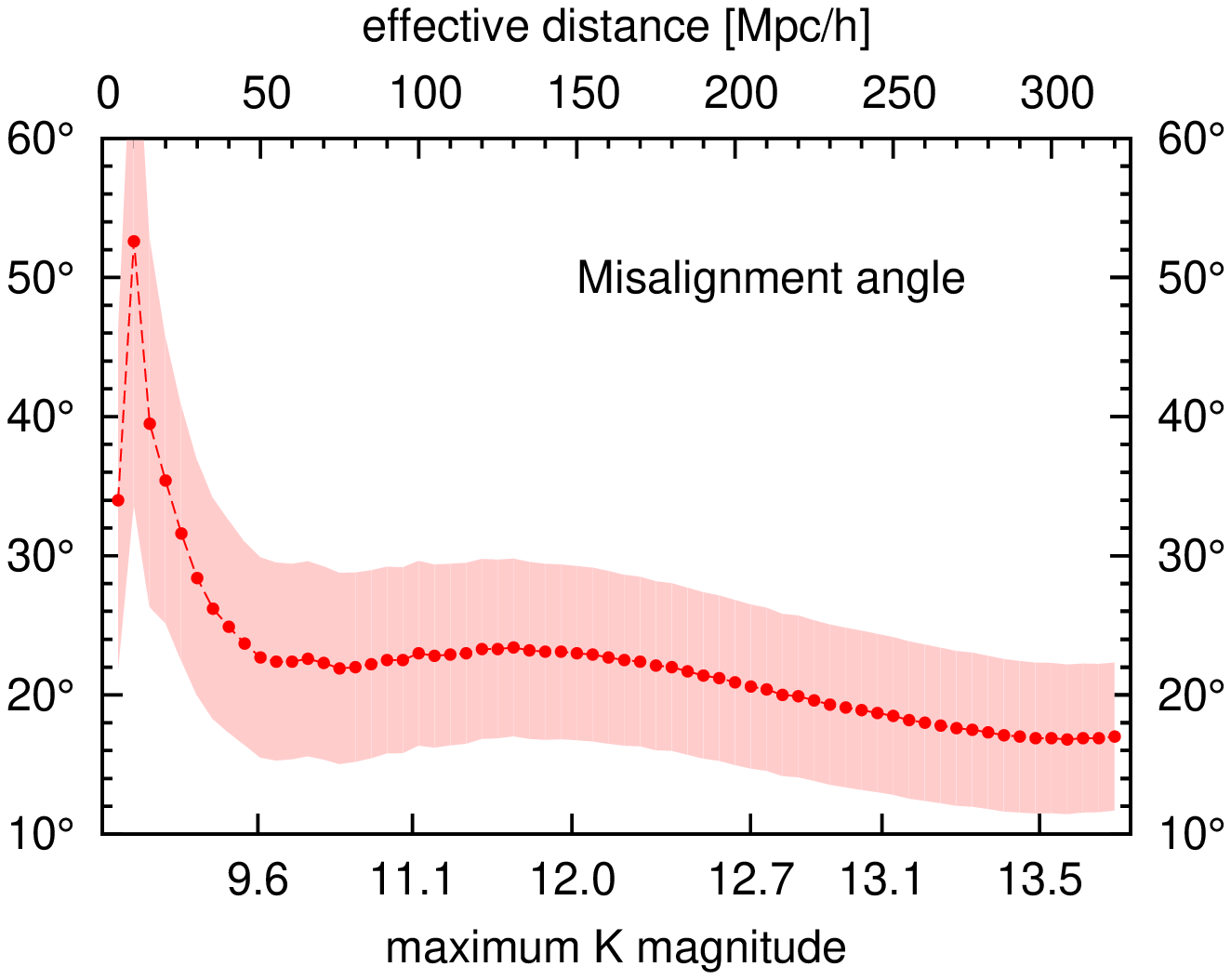,angle=-90,width=0.439\textwidth}
\caption{
Left panel: growth of the 2MASS flux dipole compared with theoretical expectations of \LCDM, the former properly rescaled. Right panel: the misalignemnt angle between the 2MASS flux dipole and LG peculiar velocity.
}
\label{Fig:growth}
\end{center}
\end{figure}

Our first method to estimate $\beta$ from the 2MASS clustering dipole was to analyze the growth of the dipole as a function of the limiting flux (maximum magnitude)\cite{BCJM11}. By comparing this growth with theoretical expectations\cite{JVW90,LKH90}, including the appropriate observational window of the 2MASS flux dipole\cite{CCBCC08}, we showed that it is consistent with the predictions of the \LCDM\ cosmological model. This consistency allowed us to estimate $\beta=0.38\pm0.04$ by appropriately rescaling the observed acceleration of the LG (left panel of Fig.~\ref{Fig:growth}).

In this analysis, to estimate $\beta$ we used the \textit{integral} dipole, where different datapoints are correlated and the error on $\beta$ obtained via the $\chi^2$ procedure is somewhat underestimated. Although we doubled the formal error, trying to include some systematics, possibly not all the effects were taken into account. For example, the \emph{misalignment angle} between LG velocity and the 2MASS clustering dipole (Fig.~\ref{Fig:growth}, right panel) was significantly larger than zero ($\sim20\degr$); however, the dependence on this angle was not included in our calculations, as it was integrated out. Our measurement, done purely in linear theory, can still be optimized. 
 
In order to obtain the most robust estimates of cosmological parameters by comparing the LG velocity and acceleration, one should preferably use the \textit{maximum likelihood} (ML) approach\cite{Strauss92}. Additionally, we need to optimize the window through which the clustering dipole is measured\cite{CCBCC08} to maximize the cross-correlation coefficient between $\vLG$ and $\gLG$\cite{CiCh04}. In case of 2MASS, this optimization in essence reduces to excluding from the calculation of the flux dipole those galaxies that are brighter than some minimum apparent magnitude\cite{CCBCC08}. This exclusion mitigates non-linear effects and shot noise from the sources located preferentially nearby. As one of the indications of this optimization, the measured misalignment angle decreases. In our calculations, we have additionally incorporated the non-linear power spectrum of velocity divergence\cite{CiCh04} and the coherence function\cite{ChCi02}. We have used state-of-the art numerical simulations to estimates these effects within the \LCDM~model\cite{H13} and found significant corrections with respect to earlier estimates. Our preliminary results of the ML method\cite{TezaMB}, which still need to be refined and will be described elsewhere\cite{BCH12}, point to a value of $\beta=0.43\pm0.03$. This result awaits further confirmation once we have included all the possible sources of errors that go into the maximum-likelihood estimate of $\beta$. At this stage, our findings provide the tightest estimates of the $\beta$ parameter from the clustering dipole to date\cite{Strauss92,Schmoldt99a,CiCh04}. They are also consistent with various determinations by other authors who have equally used 2MASS data\cite{Erdogdu06a,Davis11,NBD12GR,BDN12}.

In the near future these analyzes could be refined further with the use of an all-sky photometric redshift catalog, currently being prepared\cite{BJ13} by matching the 2MASS and WISE\cite{WISE} surveys, as well as thanks to a much deeper extended source all-sky catalog based on the latter\cite{Jar13}.

\section*{Acknowledgments}
\small{We made use of data products from the Two Micron All Sky Survey, the NASA/IPAC Extragalactic Database and
  Infrared Science Archive. We acknowledge the use of the TOPCAT software\cite{Topcat}. MB was supported by the
  Polish National Science Centre within grant no.\ N N203 509838 and by the National Research Foundation (South
  Africa). MC was supported by the Polish National Science Centre within grant no.\ 2011/01/B/ST9/06023. WAH was
  supported by the Polish National Science Centre within grant no.\ DEC-2011/01/D/ST9/01960 and by the ERC Advanced
  Investigator grant (C.~S.~Frenk) -- COSMIWAY.}

\bibliographystyle{ws-procs975x65}
\bibliography{ws-pro-MB}

\end{document}